\documentstyle[11pt,epsf]{article}

\def\beq{\begin{eqnarray}}
\def\eeq{\end{eqnarray}}
\def\bea{\begin{eqnarray*}}
\def\eea{\end{eqnarray*}}

\def\centeron#1#2{{\setbox0=\hbox{#1}\setbox1=\hbox{#2}\ifdim
\wd1>\wd0\kern.5\wd1\kern-.5\wd0\fi
\copy0\kern-.5\wd0\kern-.5\wd1\copy1\ifdim\wd0>\wd1
\kern.5\wd0\kern-.5\wd1\fi}}
\def\ltap{\;\centeron{\raise.35ex\hbox{$<$}}{\lower.65ex\hbox{$\sim$}}\;}
\def\gtap{\;\centeron{\raise.35ex\hbox{$>$}}{\lower.65ex\hbox{$\sim$}}\;}
\def\gsim{\mathrel{\gtap}}
\def\lsim{\mathrel{\ltap}}

\def\singleandthirdspaced{\baselineskip=\normalbaselineskip\multiply
    \baselineskip by 130\divide\baselineskip by 100}

\newcommand{\newc}{\newcommand}
\newc{\qbar}{{\overline q}}
\newc{\Kahler}{K\"ahler }
\newc{\deltaGS}{\delta_{\rm GS}}
\newc{\Mp}{M_{\rm Pl}}
\newc{\m}{m_{3/2}}
\newc{\N}{{\cal N}}

\begin{document}
\begin{titlepage}
\begin{flushright}
{\large hep-th/0409226 \\ SCIPP-2004/14\\WIS/23/04-SEP-DPP\\
}
\end{flushright}

\vskip 1.2cm

\begin{center}

{\LARGE\bf Hybrid Inflation and the Moduli Problem}

\vskip 1.4cm

{\large  M. Berkooz$^a$, M. Dine$^b$ and T. Volansky$^a$}
\\
\vskip 0.4cm
{\it $^a$Weizmann Institute of Science \\ Rehovot, Israel}
\\
{\it $^b$Santa Cruz Institute for Particle Physics,
     Santa Cruz CA 95064  } \\

\vskip 4pt

\vskip 1.5cm

\begin{abstract}
  We revisit some questions in supersymmetric
  hybrid inflation (SHI).  We analyze the
  amount of fine tuning required in various models, the problem of
  decay at the end of inflation and the generation of baryons after
  inflation.  We find that the most natural setting for HI is in
  supersymmetric models with non-renormalizable couplings.
  Furthermore, we argue that almost inevitably, one of the fields involved is
  a modulus, with Planck scale variation. The resulting moduli problem
  can be solved in two ways: either by a massive modulus (which
  requires some fine tuning), or an enhanced symmetry point, in which
  the moduli becomes strongly coupled to the Standard Model.  Various 
  possibilities for baryon production are discussed.


\end{abstract}

\end{center}

\vskip 1.0 cm

\end{titlepage}
\setcounter{footnote}{0} \setcounter{page}{2}
\setcounter{section}{0} \setcounter{subsection}{0}
\setcounter{subsubsection}{0}

\singleandthirdspaced

\section{Introduction}

The hypothesis that the universe underwent a period of inflation
early in its history has received striking experimental
confirmation.  From the perspective of fundamental physics, this
is surely a clue to physics at extremely high energies.  String
theory moduli would seem to be natural inflaton candidates.  If
one considers a single field, and supposes that the curvature of
the potential is everywhere of order TeV, the number of e-foldings
is naturally of order one. Sufficient inflation can arise if one is
willing to suppose a ``tuning" not worse than about 1\%.  On the
other hand, for a potential of this size, the resulting
fluctuations are far too small, of order $m_{3/2} \over M_p$.

At least two alternative proposals have been put forth to resolve this
problem.  One possibility is that some moduli are simply far more
massive than a TeV; at their minimum, supersymmetry is restored and
the cosmological constant vanishes\cite{banksinflation}.  The second
possibility is that there is more than one field involved in
inflation.  This is known as ``hybrid
inflation"\cite{Linde:1991km,Linde:1993cn}.  Guth and Randall pointed
out that hybrid inflation is particularly natural in the framework of
supersymmetry\cite{guthrandall,Randall:1996ip}, where it provides a
setting in which inflation can occur when $H\sim {\rm Tev}$.  In a
supersymmetric context, adequate inflation and a suitable fluctuation
spectrum, as we will review, can be achieved with order $1 \%$ fine
tuning\cite{lyththanks}. Guth and Randall noted that such
inflation works most naturally if some of the fields involved are flat
directions of the theory.  In this note, we will pursue the properties
of these fields further.  We will see that these fields are
necessarily moduli with Planck scale variation, as in string theory.
At generic points in the moduli space, these fields couple very weakly
to ordinary matter. As a result, in its most
appealing form, supersymmetric hybrid inflation (SHI) suffers from a
moduli problem.

In this note, we explain how and why this problem arises.  We then
discuss two widely studied solutions to the moduli problem: massive
moduli and enhanced symmetries.  Both have a rather natural
implementation in SHI.  We will argue that in SHI, the starting point
of the field evolution is probably a point of enhanced symmetry\footnote{In which case one might worry about
topological defects such as domain walls and cosmic strings.  But we
will see that the structure of hybrid inflation models naturally
avoids such difficulties.}. If
the final point is not, then the resulting moduli problem might be
cured if the modulus is quite massive, of order $100$ TeV.  This fine
tuning has been widely discussed, and is troubling.  In the context of
SHI, however, such a tuning has a possible virtue: it facilitates
achieving adequate inflation and suitable density fluctuations.  So
this is suggestive that one might be on the right track.
Alternatively, the endpoint of the moduli evolution might also be a
point of enhanced symmetry.  This may seem a contrived possibility except that in string theory, as well as many supergravity Lagrangians, there exist moduli spaces with
multiple points of enhanced symmetry.  One of these, we will see,
could explain the existence of the inflaton with its requisite
couplings to the modulus; the other could resolve the moduli problem. Solving the moduli problem with an enhanced symmetry point does not entail any fine-tuning.

In both cases, the Affleck-Dine (AD)
mechanism\cite{Affleck:1984fy,Dine:1995kz,Enqvist:2003gh,Dine:2003ax}
provides a natural way to understand the origin of baryon number, 
although other Baryogenesis scenarios can be applied.  

In the next section, we
review the hybrid inflation hypothesis and investigate how much fine
tuning is necessary to comply with observations.  We show that the
amount of fine tuning is sensitive to couplings between the inflaton
and the waterfall field and for certain non-renormalizable couplings a
mild $1\%$ tuning is sufficient.  In section 3 we explain why
moduli almost inevitably play an essential role, 
and review the
cosmological moduli problem.  In section 4 and 5, we discuss two possible
solutions to the moduli problems of these models: enhanced symmetries
and massive moduli. The four order of magnitude hierarchy (measured in
$m^2/m_{3/2}^2$) needed in the latter to obtain a moduli with
mass $100$ TeV automatically implements the $1\%$ fine tuning needed for
inflation. Section 6 contains some remarks on the efficacy of late
inflation in solving the moduli problem.  We conclude in section seven.

\section{Fine Tuning in Models of Hybrid Inflation}

Hybrid inflation in supersymmetric models,
and its virtues (and possible problems)
are best illustrated by some examples.  These are based on papers of
Guth and Randall\cite{guthrandall,Randall:1996ip}, Lyth\cite{Lyth:1999ty} and
others.  Below we begin by reviewing the idea of hybrid inflation.  We then
go on to argue that hybrid inflation suffers from a mild fine tuning
problem which can be parametrized by a single dimensionless parameter
${\cal N}$.  All other fine tuning issues are resolved if one considers
non-renormalizable couplings between $\phi$ and $\chi$.

\subsection{A Renormalizable Model}

Consider a (supersymmetric) theory with two fields, $\chi$ and
$\phi$.  $\chi$ is a modulus, by which we mean simply that it has no
potential in the absence of supersymmetry breaking.  Near the origin,
we first study a superpotential of the form:
\begin{equation}
W= \frac{1}{2}\lambda \chi \phi \phi
\label{quadratic}
\end{equation}
Note that this sort of structure is familiar in string theory.  At
points of enhanced symmetry, one typically has a variety of light
fields which couple to moduli through the superpotential or gauge
interactions.  Far away from these points, fields which can couple to
moduli will generally gain Planck scale masses, if not through
couplings of the form of eqn.  (\ref{quadratic}), then through couplings
like $\chi^n \phi \phi$.  Including soft breaking terms, the potential takes the form
(in the region of small $\chi$):
\begin{equation}
  \label{eq:9}
  V= V_o - m_\chi^2 \vert \chi
  \vert^2 + m_\phi^2 \vert \phi \vert^2 + \vert \lambda \phi \chi
  \vert^2 + \dots
\end{equation}
$V_o$ is assumed to be of order the scale typical of gravity-mediated
models, $V_o = {\cal N} m_{\chi}^2 M^2$ where $M$ is the reduced
Planck mass $M = \Mp/\sqrt{8\pi}$ and ${\cal N}$ is our fine tuning
parameter, to be evaluated below.  This corresponds to the assumption
that $\chi$ sits a distance of order $M$ from its minimum.  
At $\chi=0$, $\phi$ might also be a modulus, but this is not necessary for the mechanism
to work.  If the superpotential contains $\phi^3$ terms, for example, this will not
disrupt the mechanism described below, provided the coupling is not too large
(such terms can be prohibited by symmetries). 

The basic picture is that $\phi$ starts away from its minimum at the origin.
As long as $\lambda \phi > m_\chi$, $\chi$ is locked near the origin.
During this period, the energy is dominated by $V_o$, and inflation occurs.
Once the curvature of the $\chi$ potential becomes negative, $\chi$ rolls towards
its true minimum, and inflation ends.

A remark is now in order.  Enhanced symmetry points may involve both
continuous and discrete symmetries.  Continuous gauge symmetries don't
pose cosmological problems, so long as the U(1) of the Standard model
is not embedded in a simple group at $\chi=0$.  If it is, one has to
worry about possible production of monopoles after inflation.
Discrete symmetries, on the other hand, would seem to raise the
specter of domain walls.  If $\chi$ is zero during inflation, its
subsequent motion may break some of the discrete symmetries.  However,
this problem does not arise in this case or in the other models
we will discuss.  First, if the inflaton,
$\phi$, transforms under the discrete symmetry broken by $\chi$, the
discrete symmetry is already broken during inflation, and our
observable universe lies within a single domain. Furthermore, if we
consider our superpotential for $\chi$ and $\phi$, we must include
soft breaking effects.  Thus we expect the potential to include a term
$\m W$. This term induces a VEV for $\chi$ (of order the VEV of $\phi$), which means that $\chi$ also breaks the symmetry by a significant amount. 
One can check that the
$\chi$ VEV does not spoil the slow roll of $\phi$, or any of the other
essential features of the dynamics we describe now.  Thus there
appears to be no problem of domain walls.

One of our goals is to explore fine-tuning in this model. We will see that the observational constraints on the model are not met within the framework of
renormalizable superpotential unless one allows a large amount of fine
tuning (we are therefore led to non-renormalizable models which require much less fine tuning). The parameters that we use to quantify fine tuning are
\begin{equation}
\label{fntngparam}
{\cal N}_\chi=m_\chi^2/ \m^2,\ \ {\cal N}_\phi=m_\phi^2/ \m^2,\ \ {\cal N}=V_0/M^2m_\chi^2,\ \ \lambda
\end{equation}
With the exception of section 5, we assume that ${\cal N}_{\chi,\phi}$ are of order 1, pushing the discussion of fine tuning to the remaining parameter.

The observational constraints determine these parameter as follows. Let us first compute the number of e-foldings.  Assuming
slow roll for $\phi$, this is given by
\begin{eqnarray}
  \label{eq:6}
  N &=& \int dt H(t) = -\int d\phi \frac{V_o}{M^2 V^\prime(\phi)}
  \nonumber \\
  &=& -{\cal N}\ln(\phi_f/\phi_i).
\end{eqnarray}
So we require ${\cal N} \gsim 60$. 

Next consider the spectral index given by
\begin{equation}
  \label{eq:10}
  n_s = 1 - 3M^2\left(\frac{V^\prime}{V}\right)^2 + 2M^2\frac{V^{\prime\prime}}{V}.
\end{equation}
Since during inflation $\chi$ is heavy, one may integrate it out to be
left with an effective potential for $\phi$.  As usual, the second
term on the LHS is much smaller than the third.  One therefore
obtains,
\begin{equation}
  \label{eq:11}
  n_s \simeq 1 + \frac{2}{\cal N}
\end{equation}
which is larger than one.  Thus the more e-foldings we have, the
closer $n_s$ is to unity.  This point of view relates the smallness of
the slow-roll parameters (and therefore the scale independence of the
spectrum) to the requirement of sufficiently long inflation.  For
${\cal N} > 60$, the above agrees well with observations.

Finally, we study the size of fluctuations.   The standard relation is
\begin{equation}
  \label{eq:7}
  \frac{\delta\rho}{\rho} = \frac{V^{3/2}}{M^3V^\prime} \simeq {\cal
  N}^{3/2}\frac{m_\chi}{\phi_{60}}
\end{equation}
where $\phi_{60}$ denotes the value of $\phi$ $60$ e-foldings before
the end of inflation.  Thus one needs $\phi_{60}\gsim 10^6m_\chi$
which for the potential (\ref{quadratic}) means $\lambda \lsim
10^{-6}$.  We view this as a fine tuning which renders the renormalizable
model unattractive.

\subsection{A Non-renormalizable Model}

The problem of fine tuning can be mitigated
by considering a
non-renormalizable interaction such as\footnote{If $\phi$ is a standard model field, then it should be thought of as some linear combination
  of fields of definite charge so that these expressions are gauge
  invariant.}
\begin{equation}
  \label{eq:8}
  W = \lambda\chi\frac{\phi^n}{M^{n-2}},
\end{equation}
with the same soft SUSY breaking as in (\ref{eq:9}).
Inflation occurs in the vicinity of the value of $\phi$ where the positive mass for $\chi$ arising from $|{\partial W\over\partial \phi}|^2$ is slightly larger than the tachyonic mass of $\chi$ arising from SUSY breaking.  

The number of e-folding remains ${\cal N}$ and the spectral index also remains as in equation (\ref{eq:11}). However, the constraint from the size of fluctuations changes. At 60 e-foldings before the end of inflation
\begin{equation}
  \label{eq:1}
  \phi_{60} \simeq M\left(\frac{m_\chi}{n\lambda M}\right)^{1/(n-1)}
\end{equation}
and the constraint of $\delta\rho/\rho \simeq 5 \times 10^{-4}$ is
translated to
\begin{equation}
  \label{eq:2}
  {\cal N}^{3/2}\lambda^{1/(n-1)} \simeq 10^{-3.5+ 15(n-2)/(n-1)}.
\end{equation}
Thus for $n\ge 3$ we can set $\lambda\sim1$ and the constraint is translated to the number of
e-foldings without further fine tuning.  For $n=3$ and $\lambda$ of order one, we find ${\cal N}
\simeq 5\times 10^{2}$ which is sufficient for the required 60 e-foldings 

For larger $n$'s one requires a larger fine-tuning: $n=4\rightarrow {\cal N}\sim 10^4$ etc. As $n$ increases, the VEV of $\phi$ in which inflation occurs increases as well. As we will discuss briefly in section 3, this can be used to constrain $n$ - if
$\phi$ does not decay to Standard model fields before matter-radiation equality
(which implies either large enough mass, or large enough couplings to the SM),
then $n\ge 5$ is excluded. 
 

In \cite{guthrandall}, it was necessary to impose a hierarchy between
$m_\phi$, $\m$ and $m_\chi$.  Since $m_\phi$ and $m_\chi$ are both
expected to be proportional to $\m$ one can view this as fine tuning
of ${\cal N}_{\chi,\phi}$.  The need for this hierarchy originate from
the evolution of $\chi$.  The authors show that unless inflation ends
sufficiently rapidly, $\chi$ creates unacceptably large density
fluctuations.  The requirement of rapid ending requires $m_\phi \ll \m
\ll m_\chi$.  As we now argue, this hierarchy is not necessary when
one takes into account all of the expected soft breaking terms.

Indeed, given the superpotential (\ref{eq:8}), it is inevitable that
soft breaking terms of the form $\m W +$ c.c. will appear at low
energy after supersymmetry breaking. Such tadpole terms trigger a VEV
for $\chi$ of order $\phi$ (assuming $\lambda = {\cal O}(1)$)
already during inflation.  Fluctuations around this central value are
of order\footnote{Here one assumes $\chi$ is moving fast enough to catch up
with its minimum.  This is true since $\dot \chi = (2-n)\dot \phi$ and the
minimum depends on $\phi$.}  $H$ and are therefore small ($H/\chi \simeq  (\m/M)^{n-2\over
n-1} \ll 1$ for $n\geq 3$).  Thus one expects the fluctuations of $\chi$ to
be of the same order as the fluctuations of $\phi$ both of which agree
with observations.  This differs
considerably from the situation where $\chi$ has no VEV and
fluctuations may be large (which is the case in \cite{guthrandall}).

\section{The Moduli Problem in Hybrid Inflation}

While it is not always clearly stated, the assumptions of the previous
section almost inevitably imply that $\chi$ is a modulus.  By modulus,
here, we mean a field with no potential in the limit of vanishing
supersymmetry breaking.  As a prototype, suppose there were terms in
the superpotential of the form \beq W_{\chi} = {\chi^{m+3} \over M^m}.
\label{wchi}
\eeq Suppose, again, that $\chi=0$ is the region in which inflation
occurs, and at this point $\chi$ has soft terms, $-\m^2 \vert \chi
\vert^2 + \m W$. So the minimum of the $\chi$ potential is at a scale less
than $M$, \beq \chi \approx M \left(\frac{\m}{M}\right)^{1 \over m+1}.
\eeq The energy shift between the symmetric minimum and this final
minimum is then much less than $\m^2M^2$: \beq \Delta V \approx
m_{3/2}^2 M^2 \left ( {m_{3/2}^2 \over M^2} \right ) ^{1 \over m+1}.
\eeq

It seems difficult to build models of inflation with modest values of
$m$, at least with the usual assumptions about the scales of
supersymmetry breaking.  For example, plugging in eqn. (\ref{eq:6}),
the number of $e$-foldings will be suppressed by \beq \epsilon = \left
  ( {m_{3/2}^2 \over M^2} \right ) ^{1 \over m+1}.  \eeq

In \cite{guthrandall}, flat directions of the MSSM are discussed
extensively.  But for all of these, the gauge symmetries alone are not
enough to prevent corrections of the type of eqn. (\ref{wchi}).
Additional symmetries {\it can} render these directions exactly flat.
Discrete R symmetries can forbid infinite numbers of operators and
yield such flat directions\cite{dineseiberg}.  Moreover, in string
theory, it is well known that such flat directions can sometimes arise
by what appears to be, from a low energy point of view, an accident.
The absence of such terms might have implications for proton decay.

The dynamics of the system after inflation is the interplay of two
points - $\chi=0$ where inflation occurs, and the true minimum of the
system, which is a distinct point to which $\chi$ rolls after
inflation. One expects $\chi=0$ to be a point of enhanced symmetry. The simplest reason is that enhanced discrete symmetries are needed to obtain a potential which is suitable for inflation. 
A related argument is that it is important that near the origin,
$\chi$ couples to the
light field $\phi$ in the superpotential. This is most natural if the
point $\chi=0$ is a point of enhanced symmetry, and the light $\phi$ is
associated with this enhanced symmetry.

Given that $\chi$ is a modulus, and that around the true minimum
its couplings to Standard Model fields are likely to be Planck
suppressed, it potentially suffers from the standard moduli problem of
string cosmology\cite{moduliproblem}.  Once $\phi$ is small enough,
the modulus, $\chi$, begins to roll towards its minimum and inflation
ends.  Away from $\chi=0$, the potential for $\chi$ is assumed to take
the form: \beq V(\chi)= m_{3/2}^2 M^2 f(\chi/M) \eeq $\chi$ reaches
its minimum in roughly a Hubble time, and begins to oscillate.  It
behaves like pressureless dust, with $p=0$, and dominates the energy
density of the universe until it decays.  It is usually assumed that
the width of such a particle is of order \beq \Gamma= {m_\chi^3 \over
  M^2} = 10^{-27}~ {\rm GeV}.  \eeq This corresponds to a reheat
temperature of order $10~{\rm KeV}$.  This spoils the successes of the
standard cosmology.

Reference \cite{guthrandall} considered the possibility that $\chi$ is
one of the approximate flat directions of the MSSM.
The authors did not address the question of
the required flatness of the $\chi$ potential.  As we indicated, the
potential must be extremely flat; this might be accounted for by
symmetries or by some stringy phenomenon.
We have suggested that the 
lightness of $\phi$ during inflation might result from
an enhanced symmetry at this point
in the $\chi$ potential.  As we will discuss later, in string theory
it sometimes occurs that a moduli space has multiple enhanced symmetry
points.  If this accounts for the features of $\chi$, then the $\chi$
couplings to the SM need not be suppressed at the minimum, and, as we
will shortly explain, there would be no moduli problem.

It should be noted that $\phi$ might have a moduli problem of its own.
As $n$ increases in equation (8), the VEV of $\phi$ in which inflation
occurs is pushed to higher values, and more energy is stored in
$\phi$. If $n>4$ and $\phi$ is coupled to the SM only with
$M$-suppressed couplings, it will have its own moduli problem,
irrespectively of how the $\chi$ moduli problem is resolved. The
solutions to the $\phi$ moduli problem are either the ones that we
will discuss for $\chi$ (with additional fine tuning if $\phi$ is a
massive moduli), or one concludes that $n<5$.

A final remark is in order.  In the following two sections we present
two possible solutions to the cosmological moduli problem.  Both
solutions involve a long epoch in which the universe is matter
dominated before $\chi$ finally decays. During this time, density
fluctuations grow and one may worry of an excess density of primordial
black holes and/or Q-ball formation. Moreover, in the enhanced symmetry scenario, a
positive or negative pressure might build up due to running of the
mass parameter\cite{Enqvist:2003gh}.  The former could prevent the
formation of these objects while the latter would allow Q-ball generation; these
Q-balls could act as dark matter and create the baryon
asymmetry\cite{Enqvist:1998en}.  In either case, this would constrain
the low energy theory considerably.  We will return to this point in
future work\cite{Berkooz:future}.

\section{Enhanced Symmetries}

Since $\chi$ couples to massless fields, it is natural to suppose that
$\chi=0$ is a point of enhanced symmetry (although there could be
points in the moduli space where, accidentally, some fields are
massless). Such an enhanced symmetry is necessary for the structure
hybrid inflation potential. This also makes it natural that the $\chi$
field starts near the origin.  Suppose, however, that the true minimum
of the $\chi$ potential lies at another point of enhanced symmetry.
Then there is no problem with this modulus; its lifetime is of order
$m_{3/2}$ (modulo factors of coupling constant).  In addition, $\chi$
might be some linear combination of fields of the MSSM, i.e. a flat
direction of the MSSM.  So motion in this direction could lead to
production of baryons.  Indeed, one might worry that unless CP
violation is very small, there will be too many baryons.  We will see
that this is the case.

It is easy to see how discrete $R$ symmetries can make some of the
flat directions of the MSSM exact.  Suppose, for example, that one has
a $Z_N$ symmetry ($N>3$), under which the Higgs transform with phase
$\alpha= e^{2\pi i/N}$, while the quarks and leptons are neutral.
Suppose that the superpotential, $W$, transforms with phase $\alpha$.
In addition, suppose that there is a $Z_2$ symmetry (non-R) under
which the quarks and leptons are odd, and all other fields are even.
Then the usual Yukawa couplings are all allowed.  (The $\mu$ term
vanishes as well; when supersymmetry is broken, the $R$ symmetry is
broken as well, and a $\mu$ and $B_\mu$ terms may be generated). But
terms in the superpotential involving powers of the various quark and
lepton invariants, $\bar u \bar d \bar d$, $Q Q Q L$, $\bar u \bar u
\bar d \bar e$, etc., are all forbidden.  Invariants involving Higgs
fields are permitted, but typically must involve several fields.  It
is easy to see that there are numerous exact flat directions: $Q Q Q
L$, $\bar u \bar d \bar d$, etc.

One might object that these symmetries have anomalies.  Since in an
underlying theory, one expects any discrete symmetries are gauge
symmetries, this might seem inconsistent.  But it is not difficult to
anomaly free variants with many flat directions.  For example, if the
third generation leptons, $L_3$ and $\bar e_3$ transform with phase
$\alpha^{-2}$ and $\alpha^2$, respectively, there are no anomalies,
yet one can check that there are many flat directions.  Interestingly,
in this case, one can check as well that all baryon violating dimension four
and five operators are forbidden.  To account for neutrino mass, it
may be necessary to modify these assignments, but this example
illustrates that anomalies are not an obstacle.

One might also ask: how plausible is it to find more than one point of
enhanced symmetry in a moduli space.  Certainly, in toroidal
compactifications of weakly coupled string theories, there are many
points of enhanced symmetry at which {\it all moduli but the dilaton}
are charged.  E.g.  for the heterotic string on $T^6$, there are
points of $SU(2)^6$ and $SU(3)^3$ symmetry.  In addition, at these
points, one has unbroken discrete symmetries.  It is easy to check
that all moduli but the dilaton transform under one or more of these
symmetries.

\subsection{Reheating and Barygenesis}
\subsubsection{Moduli Space of Real Co-Dimension$ > 1$}

In the enhanced symmetry scenario the field $\chi$ becomes part of the
SM near its true minima. We will see that the cosmological evolution of
the model does not depend sensitively
on where $\chi$ sits in the SM.
The features of this field that we use are
\begin{enumerate}
\item In general, it could be that the inflationary point in field
  space is characterized by several moduli that are fixed during
  inflation. We will denote these by $\rho_i,\ i=1..N-1$. As we move to
  large values of $\chi$, it mixes with $\rho_i$ and they become
  indistinguishable. We will denote the entire collection by $\rho_i,\ 
  i=1..N$. The true minima is assumed to be an isolated point in this
  $N$ dimensional space.
\item The fields $\rho_i$ becomes strongly coupled to the SM only
  within a circle of size $m_{\rho_i}\sim m_{3/2}$ in this moduli
  space. At large $\rho$, the fields to which these moduli couple have
  mass of order $\rho$; interactions of the $\rho$ fields with
  remaining light fields are suppressed by factors\footnote{If $\delta
    \rho$ is the canonically normalized fluctuation of the field, and
    $\rho_0$ is the VEV, then the interaction to the SM is through
    $(\delta\rho/\rho_0)$, i.e., ${\cal L}=... + log(\rho)*{\cal
      O}_{(SM)}$ } of $\rho^{-1}$.
\item Once the amplitude of the field $\rho_i$ becomes small
enough, these fields decay.  Equating decay rates and expansion
rates, this typically occurs when the amplitude is of order 
$10^6-10^7$ GeV. The reheat temperature is of order $10^2 m_\rho\sim 10^5$ GeV.
\end{enumerate} 
For now we will keep the distinction between $\phi$ and the $\rho_i$.

When working in the supersymmetric framework, the fields $\rho_i$
parameterize a complex manifold and the real co-dimensionality of true
minima is necessarily greater than one (still, the co-dimension
1 case is interesting, and we will discuss it in the next
subsection).
For this larger co-dimension case, it is straightforward to see that the
reheat temperature is typically two or three
orders of magnitude above $m_{3/2}$.  The decay rate of the
$\rho_i$ fields is $\Gamma \approx {m^3\alpha^2 \over \rho^2}.$
This leads to a reheat temperature of order
\beq
T_{rh} \approx m (\alpha^2 M_p/m)^{1/6}.
\eeq
and quite
a large baryon number
\beq
n_B/s\sim 10^{-2} \times ~{\rm CP-phases} .
\eeq

To understand this estimate, suppose that the
$\rho_i$ fields have baryon number $q_i$.
Parameterizing, $\rho_i = |\rho_i|e^{i\theta_i}$, the baryon number is
then given by
\begin{equation}
  \label{eq:24}
  n_B = \Sigma_i q_i \int d^3x \left(\rho_i^*\dot\rho_i-\dot\rho_i^*\rho_i\right) \simeq
  \Sigma_i q_i\int d^3x|\rho_i|^2\dot\theta_i
\end{equation}
which leads to $n_B/s$ of the order,
\begin{equation}
  \label{eq:25}
  \frac{n_B}{s} \simeq
  \left(\Sigma_iq_i|\rho_i|^2\dot\theta_i\right)\frac{T_R}{E_\rho}
\end{equation}
where $E_\rho$ is the total energy in $\rho_i$.
At the time of decay, the above becomes $n_B/s \simeq
\left(\Sigma_iq_i\dot\theta_i\right)/\m (10^{-2})$. Here $\dot\theta_i$ depend
on the baryon violating interactions. For example the Kahler
potential is likely to break the $U(1)_B$ through terms such as
\begin{equation}
  \label{eq:26}
  K = \rho_i\bar\rho_i\left(1 + \frac{\rho_i^n}{M^n}+
  \frac{\bar\rho_i^n}{M^n}+...\right)
  + ...
\end{equation}
Immediately after inflation, when $\rho_i \sim M$, the potential
is such that the $\rho_i$ phase gets a ``kick", generating some
$\dot\theta_i \sim \m$.  Thus unless one tunes the CP and baryon
violating terms to get $\dot\theta_i/\m \ll 1$ at the end of
inflation, one ends up with order one baryon number. Adequately
suppressing the baryon number seems to require tuning, so this
does not seem a promising mechanism.

One may hope that the inflaton field itself may serve as the AD
field. However, checking the details of the model shows that this is
not the case. $\phi$ will produce the largest amount of baryon number under two
conditions: 1) it decays to SM only when $\rho_i$ decays and not
earlier and 2) there are terms in the superpotential which produce a
non-zero $\dot\theta$.   However, even if both conditions are
fulfilled, since inflation ends when $\phi/M$ is
small (eq. (\ref{eq:1})), one obtains $E_\phi/E_\chi \ll 1$ and an
extremely eccentric orbit with $\dot\theta/\m \ll 1$ both of which
suppresses the baryon number below the observed value.

If we are not willing to accept fine tuning, we can enumerate other 
possibilities in the case that the moduli are not weakly coupled
in their ground state, and the reheat temperature is $m_{3/2}$ or larger:

\begin{enumerate}
\item An additional SM field flat direction which carries baryon number
and is not a modulus has a VEV. Using discrete symmetries one can then
arrange the appropriate Baryon number via an AD mechanism\cite{Dine:1995kz,Dine:2003ax}.

\item Other possibilities might include EW baryogenesis.  Although the
usual SM-EW baryogenesis fails to generate a sufficient amount of
baryons, many extensions to the SM work fine.  The simplest, is the
MSSM-EW baryogenesis which has more CP phases and which enhances the
strength of the phase
transition\cite{Carena:1996wj,Farrar:1996cp,deCarlos:1997ru,Laine:2000rm}.
Other examples of extensions of the SM or the MSSM exist (see for
example \cite{Berkooz:2004kx,Menon:2004wv}).

\item Leptogenesis at the TeV scale can be ignited\cite{Hambye:2004jf}.
\end{enumerate}

\subsubsection{Moduli Space of Co-dimension=1}

Although less interesting in the supersymmetric context\footnote{The
  co-dimension 1 dynamics is relevant for the supersymmetric case if
  the motion of the supersymmetric complex field around its minima is
  highly eccentric. This can be achieved but requires fine tuning.},
the case of co-dimension one is special since the reheat temperature
is significantly higher.  The physical reason for this is that a one
dimensional field must go through the regime of strong coupling in
each oscillation and thereby the release of energy begins much sooner.
Since the field is real, it does not carry baryon number.

In this section we will use a yet simpler model for the coupling of $\chi$ to the SM. $\chi$ is now a real field. 
The assumption is that there is a real modulus $\chi$ with mass $m_{3/2}$
that starts at a VEV of order $M$.  Around VEV zero of the field there
is small region of size $\m$ in field space in which the field is
strongly interacting with the SM, and the decay rate there is
$\Gamma\sim \m$ in the sense that in flat space the loss of energy is
\begin{equation}
  \label{eq:29}
  \partial_t E_\chi=-\Gamma E_\chi.
\end{equation}

We wish to compute the reheat temperature.  The EOM for the field
reads,
\begin{equation}
  \label{eq:3}
  \ddot\chi + 3H\dot\chi + \theta(\m^2-\chi^2)\Gamma\dot\chi_i + \m^2\chi_i = 0.
\end{equation}
Multiplying the above by $\dot\chi$ and averaging over one cycle
we find 
\begin{eqnarray}
  \label{eq:30}
  &&\partial_t E_\chi+3HE_\chi+m^2_{3/2}\Gamma \sqrt{E_\chi} = 0,
  \\
  &&\partial_t E_{\rm SM}+4HE_{\rm SM}-m^2_{3/2}\Gamma \sqrt{E_\chi} = 0,
\end{eqnarray}
where we've defined $E_\chi$ ($E_{SM}$) to be the energy in $\chi$
(SM).


Assuming that the energy is dominated by $\chi$ we can solve the 1st
equation. Changing variables to $a=e^\eta$, the equation becomes
\begin{equation}
  \label{eq:31}
  \partial_\eta (e^{3\eta}E_\chi)=-\sqrt{3}e^{3\eta}\m^2\Gamma M
\end{equation}
and the solution is
\begin{equation}
  \label{eq:32}
  E_\chi=e^{-3\eta}m_{3/2}^2M^2 - \left(1-e^{-3\eta}\right){1\over \sqrt{3}} \m^2\Gamma M.
\end{equation}
The shift to a negative asymptotic number is the advantage relative to
the standard moduli issue. 
This energy goes to zero when the scale factor is increased by
roughly
\begin{equation}
  \label{eq:33}
  a/a_0 \simeq (M/\Gamma)^{1/3}.
\end{equation}
Equation (\ref{eq:32}) is to be trusted
only until $E_\chi=E_{SM}$, but this occurs slightly before (\ref{eq:33}).

Plugging this into the equation for the energy in the SM we get the
equation (again in the epoch in which $E_\chi$ dominates)
\begin{equation}
  \label{eq:34}
  \partial_\eta  E_{\rm SM} + 4E_{\rm SM} - \sqrt{3}\m^2\Gamma M
  = 0,
\end{equation}
and the solution (with $E_{SM}=0$ as initial condition) is
\begin{equation}
  \label{eq:35}
  E_{\rm SM}={\sqrt{3}\over4}\left(1-e^{-4\eta}\right)\m^2\Gamma M.
\end{equation}

One can evaluate when the energy in the SM and in $\chi$ are
approximately the same. This turns out to be very close to the point
in which the amplitude of $\chi$ is $\m$. Thus when they are equal the
field $\chi$ is already strongly interacting and there is no need to
discuss the SM dominated universe with a distinguished $\chi$
component.  Finally, the reheat temperature is approximately
$\left(\m^2\Gamma M\right)^{1/4}$ or for $\Gamma\sim \m$, $T_{\rm Re}
\simeq \left(\m^3M\right)^{1/4} \simeq 10^7$ GeV.  Such reheat
temperature cannot ignite the usual leptogenesis which requires
$T_{\rm Re} > 10^8-10^{10}$ GeV
\cite{Davidson:2002qv,Buchmuller:2002jk,Giudice:2003jh}.  Other
possibilities are (non-supersymmetric) low energy variants of
EW-baryogenesis\cite{Berkooz:2004kx} and possibly low-scale
leptogenesis\cite{Hambye:2004jf,Grossman:2003jv,D'Ambrosio:2003wy} (although these models mostly work in the supersymmetric
framework and are therefore less relevant).

\section{Massive moduli}
\label{Massive moduli}

As has often been noted\cite{moduliproblem}, a modulus with mass of order $100$ TeV will,
when it decays, reheat the universe to temperatures of order $10$ MeV,
restarting nucleosynthesis.  This large mass is usually felt to
represent a disturbing fine tuning.  But in the context of slow roll
inflation, as presently understood, some fine tuning appears
inevitable.  Although as we now argue the fine tuning required to
solve the moduli problem is slightly more significant, it is still
interesting that this fine tuning of the mass can simultaneously solve
the moduli problem and yield a suitable number of $e$-foldings and a
satisfactory fluctuation spectrum, without further fine tunings.

To be more precise, let us restore our three fine tuning parameters
$\N_\phi$, $\N_\chi$ and $\N$.  The moduli problem requires $\N_\chi
\simeq 10^4$ and we require no further fine tuning in $V_0$, namely,
$\N = 1$.   With this, the observables from inflation take the form,
\begin{eqnarray}
  \label{eq:27}
  N &=& \frac{\N_\chi}{\N_\phi}
  \\
  n_s &=& 1 + \frac{2\N_\phi}{\N_\chi},
\end{eqnarray}
and
\begin{equation}
  \label{eq:28}
  \frac{\delta\rho}{\rho} = \frac{\N_\chi}{\N_\phi}\left(\frac{m_\chi}{M}\right)^{n-2/n-1}
\end{equation}
So for $n=3$ one requires $m_\phi$ to have a mass only slightly above
$\m$, namely $\N_\phi \simeq 10$, in order to explain the fluctuation
spectrum.  For $n=4$, $m_\phi$ must be anomalously small with $\N_\phi
\simeq 10^{-1}$.

In the heavy modulus scenario, baryons can be produced through
modulus decays, but this requires significant violation of $R$-parity,
which may be problematic.  An alternative possibility is AD baryogenesis,
which is quite natural in such a framework.
This requires that, in addition to the heavy moduli field(s) $\chi$, that some
of the SM flat directions be {\it very} (possibly exactly)
flat.  Call this (these) directions $\Phi$. For simplicity,
suppose that it is exactly
flat.  
%
It is possible -- and natural -- that, when $\chi=0$,
the minimum of the $\Phi$ potential is far from
the origin.
When
when $\chi$ begins to roll towards its minimum, so does $\Phi$.  A
baryon asymmetry will be produced in the motion of $\Phi$.  If the $\Phi$
potential is flat, and the initial
$\Phi$ amplitude is of order $M_p$, this will be of order one per $\Phi$ particle, times
CP violating phases.  Note that the number of $\Phi$ particles is comparable to the number
of $\chi$ particles.

Initially, there will be a period where $\Phi$ and $\chi$ are oscillating simultaneously.
But near the origin, $\Phi$ couples to light fields, so it will decay earlier.
The baryon number can be estimated by noting that, just as there is of order one
baryon per $\Phi$ particle, there is of order one baryon per $\chi$.
Since the reheating temperature is of order $10^{-2}$ GeV, the number of photons per
$\chi$ particle is $10^{5}/10^{-2} = 10^7$.  So the baryon per photon ratio is of
order $10^{-7}$ times CP violating phases.  So this is quite naturally in a reasonable
ballpark.
If $\Phi$ is an approximate modulus, it can still generate a suitable baryon number,
provided that its potential is flat enough.  
The $\Phi$ decays can also lead to production of a suitable dark matter
density\cite{Moroi:1999zb}.  A scenario which realizes these possibilities
has been discussed in \cite{Dine:2004is}.

\section{Weak Scale Inflation?}

Weak scale inflation is often mentioned as a solution of the moduli
problem, and one might try to invoke it here in order to understand solve the
moduli problem in this context.  Indeed, we could consider a theory with an
extra field, $\Phi$.  If we tune the potential of the $\Phi$ field suitably,
we could obtain additional inflation, now with very small fluctuations.
As described by Randall and Thomas\cite{Randall:fr}, this number should not be larger than
about $25$, so as not to spoil the successes of the earlier stage of inflation.

In our scenario, however, additional weak scale inflation (which is in fact the
scale of hybrid inflation) is not a possible solution since solving
the moduli problem for $\chi$, introduces a new moduli problem for the
weak scale inflaton.  For that reason, weak scale inflation at its
best, still leaves a modulus.  

In fact, it is not widely appreciated that {\it weak scale inflation
quite generally does
  not generally solve the moduli problem.}  The difficulty is that, even when $H
\sim M_p$, the minimum of the moduli potential has no reason, in
general, to coincide with its $H=0$ minimum.  Corrections to the
moduli potential will include terms such as \beq H^2 \vert \chi
\vert^2 \eeq One might hope that, during inflation, the modulus will
be driven to the instantaneous minimum of its potential, and then will
remain in the instantaneous minimum.  But it is easy to see that this
is not the case.  Call $V(\chi;t)$ the time-dependent potential for
$\chi$, and $\chi_o$ the instantaneous minimum.  Then writing $\chi =
\chi_o + \delta \chi$, the equation of motion: \beq \ddot \chi + 3 H
\dot \chi + V^\prime(\chi,t) =0 \eeq becomes \beq \ddot \delta \chi +
3 H \dot \delta \chi + V^{\prime \prime}(\chi_o;t) = - (\ddot \chi_o +
3 H \dot \chi_o).  \eeq The problem is that, when $H$ is comparable to
$m_{3/2}$, all of the derivatives are of the same order, and the
source term leads to a large $\delta \chi$.  Note that we need a {\it
  huge} suppression, by many orders of magnitude, of $\delta
\chi$.

\section{Conclusions}
In this paper, we have revisited various issues in hybrid inflation.
In particular, we examined how finely tuned the model must be; does it
exhibit a moduli problem and how can this problem be solved; and what
is the natural way to produce baryons after inflation.

We have argued that the amount of fine tuning is highly sensitive to
the couplings between the inflaton $\phi$ and the waterfall field
$\chi$.  We found that for a certain non-renormalizable couplings $W =
\chi\phi^3/M$ there exist a single mildly tuned parameter ${\cal N}
\gsim 5\times 10^{2}$ which then allows for a sufficiently long
inflation together with a correct spectral index and density
perturbations.  Taking into account also soft breaking terms, we
showed that no further tuning is necessary to prevent $\chi$
from creating large fluctuations at the end of inflation.
From this point of view, one relates the
large number of e-foldings to the scale invariance of the spectrum.

Next we argued that $\chi$ is necessarily a modulus and therefore
it decays late, creating a moduli problem after inflation.  Two
solutions were offered.  The first involves enhanced symmetry points
where one assumes that the moduli space consists of an inflationary
point and a SM point both of which have enhanced symmetries.  This in
particular means that $\chi$ is strongly coupled near the SM point
and therefore has no moduli problem.
Since it is most natural to work in the supersymmetric framework, the
reheat temperature turns out to be quite low, $T_{\rm Re} \sim 10^5$
GeV.   Furthermore, soft breaking terms ensure that domain walls are
created after SUSY breaking and before inflation and therefore impose
no cosmological problem. 

The second possible solution is heavy moduli.  Usually, this
possibility seems disturbingly fine-tuned, but
in the case of hybrid inflation
this same tuning can explain the number of e-foldings
and the size of the fluctuation spectrum.
It is tempting to imagine that the tuning might have an anthropic
explanation.  This might fit, for example, into a scenario in which
the landscape predicts supersymmetry, put forward in \cite{Dine:2004is}.

With such low reheat temperature, it is more difficult to create 
the baryon asymmetry.  We argued that the most natural way is 
through the AD mechanism.  However, in both cases we showed that 
$\chi$ cannot be the AD field and one must introduce another flat 
direction which carries baryon number.

\section{ Acknowledgments }

We would like to thank Anthony Aguirre, Tom Banks, Yossi Nir and Yael
Shadmi for useful discussions. The work of MB is supported by the
Israeli Academy of Science centers of of excellence program, by the
Minerva Foundation, by EEC-RTN-2000-001122, by the Einstein center for
theoretical physics, and by the Harold Blumenstein Foundation.

\end{document}